\documentclass{apex}

\title{
    Optical measurement of acoustic radiation pressure of the near-field
	acoustic levitation through transparent object
}

\author{
    SATOSHI Nakamura, TOSHIAKI Furusawa, YASUHIRO Sasao, 
	KOGURE Katsura, KONDO Naoki (Teikyo-Univ.)
}

\inst{
    Teikyo-univ, 1-1 Toyosatodai, Utsunomiya, Tochigi, 320-8551, Japan
}

\abst{
	It is known that macroscopic objects can be levitated for few to several hundred micrometers by near-field acoustic field and this phenomenon is called near-field acoustic levitation (NFAL).
	
	Although there are various experiments conducted to measure integrated acoustic pressure on the object surface, up to now there was no direct method to measure pressure distribution. In this study we measured the acoustic radiation pressure of the near-field acoustic levitation via pressure-sensitive paint.

}

\begin{document}
\maketitle

	Force that pushes the object placed in the sound field away from the sound source has been known for a long time. This force is referred to as acoustic radiation pressure.
	
	This force is weak when object is in the wave's far field. But it is later discovered that if the object is put close the sound source, typically in a distance of few to hundred micrometers, object of more than several grams can be levitated. This phenomenon is referred to as near-field acoustic levitation.
	
	Since the discovery, various experimental efforts and theoretical considerations have been devoted to explain this phenomenon. But the closeness of the object surface and the sound source, which is the very cause of strong levitation pressure, prevented the direct measurement of pressure distribution.
	
	In this paper we show our new experimental technique and its result that can directly measure the pressure distribution on levitated object's surface by observing the luminescence of pressure-sensitive paint through transparent object.

	Pressure-Sensitive Paints (PSP) are paints for measuring air pressure or local oxygen concentration, which are usually used in the field of aerodynamics. They are usually comprised of phosphorent dyes and binder. The latter is for fixing dyes  on object surface.

	When this paint is illuminated with light of a specific wavelength (excitation light), the dye molecule gets pumped up to its excited state and emits a photon of specific wavelength (phosphorescence) and returns to the ground state.
	
	During this process oxygen also causes non-radiant relaxation, which works as phosphorescence quenching.
	
	Since oxygen concentration correlates with the atmospheric pressure it is possible to measure the pressure near the coated surface from the intensity of the phosphorescence.
	
	Relation between pressure and luminance is well described by Stern-Volmer equation.(exp.1)

	\begin{equation}
		\frac{ I_{ref} }{ I } = A(T) + B(T) \cdot \frac{ P }{ P_{ref} }
	\end{equation}
	
	From expression 1, we can see that it is important to correctly evaluate the effect of temperature because the intensity of the phosphorescence depends both on the oxygen concentration and the temperature.
	
	Time response of PSP is determined by how fast oxygen can penetrate the binder to reach the dye molecules.

	We employed as binder the thin-layer chromatography(TLC) plate following the experiment of McGraw et al.
	
	Figure 1 is a diagram of the measurement system.
	
	Ultraviolet light for PSP excitation is placed above the TLC plate which is  initially put on the horn tip. Phosphorescence intensity with and without the horn vibration is measured by high speed camera.
	
	The horn was designed to oscillate at 38 kHz. Its diameter is Φ30mm.
	
	The model of high-speed camera was FASTCAM made SA5 model 1300K-M1. UV light source was UltraFire WF-502B which consist of UV LEDs of 395-410nm peak wavelength. We inserted to exclude this excitation light from measurement by placing an ultraviolet absorbing filter(FUJIFILM made SC40) in front of the lens of the camera.
	
	We acquired 100 images in each single measurement sequence with camera conditions of 60fps, 12bit depth gray scale, image resolution 512x512, and then integrated to to form a single image. 
	
	Usually, Pressure sensitive paint used in the pressure distribution measurement of surfaces, such as aircraft model in wind tunnel testing. The experiments using pressure sensitive paint on the detection of the sound pressure was performed by M.McGraw et al.
	
	In this experiment is applied to bind created in the following recipe a PSP on the thin-layer chromatography plate (TLC). This recipe is following the experiment of this McGraw et al.
\indent
	\begin{tabular}{lll} 
	pigment & : & PtTFPP, 3mg\\
	binder  & : & MOMENTIVE made RTV 118, 65mg\\
	solvent & : & Dichloromethane\\
	TLC     & : & Sigma-Aldrich made Z122777-25EA\\
	\end{tabular}
	
\begin{figure}[h]
\begin{center}
	\includegraphics[width=12cm]{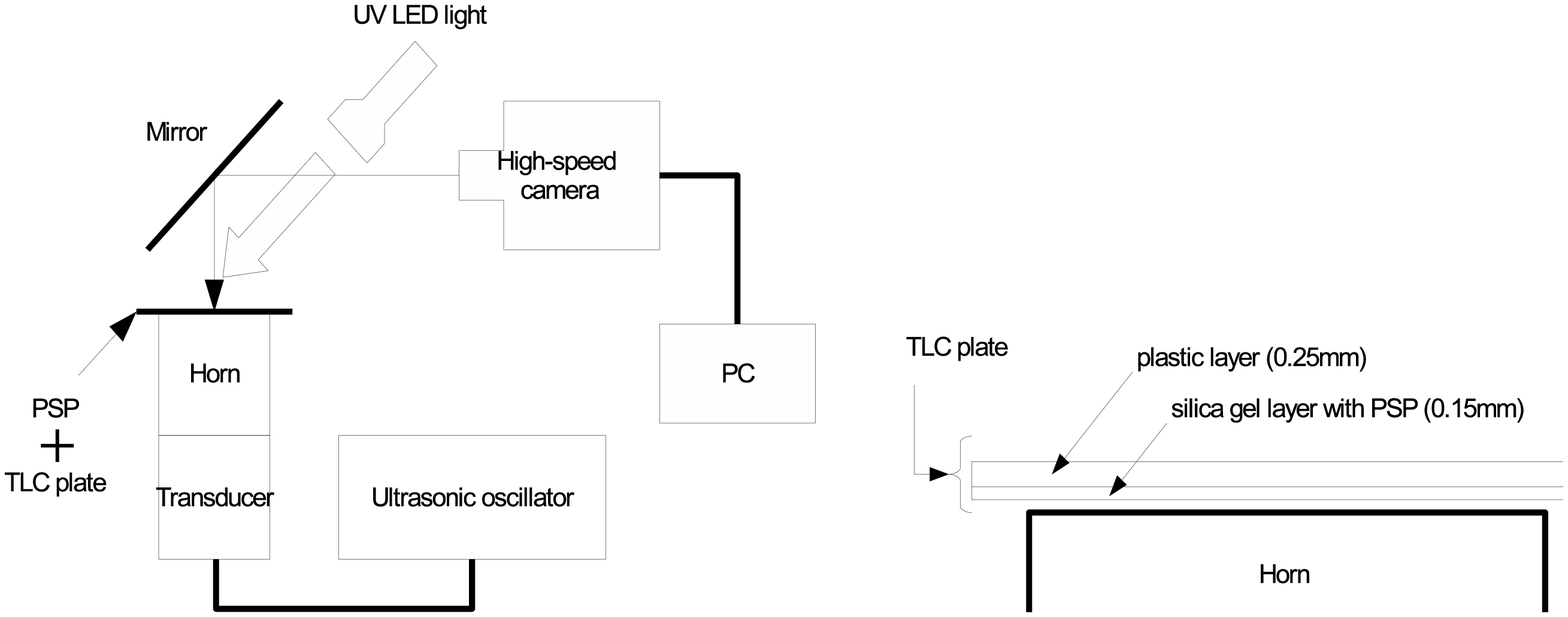}\\
	Fig.1. Diagram of the mesuarement system
\end{center}
\end{figure}

	After obtaining experimental results, we performed image processing on the acquired images and visualized the change of phosphorescence intensity. Program was coded in R language.
	
	From integrated images "with vibration" and "without vibration" we calculated the rate of change of phosphorescence intensity according to equation 2.

	\begin{equation}
		\Delta I_r = \frac{ I_{ref} - I } { I_{ref} }
	\end{equation}
	
	Figure 2 is the contour plot of the obtained phosphorescence intensity change.

	Also to see the rate change more clearly, we show a cross-sectional view of a contour plot. This plot is obtained from integration of one-dimensional-like strip area along the line crossing the center of horn (Figure 3)
	
\begin{figure}[h]
\begin{center}
	\includegraphics[width=10cm]{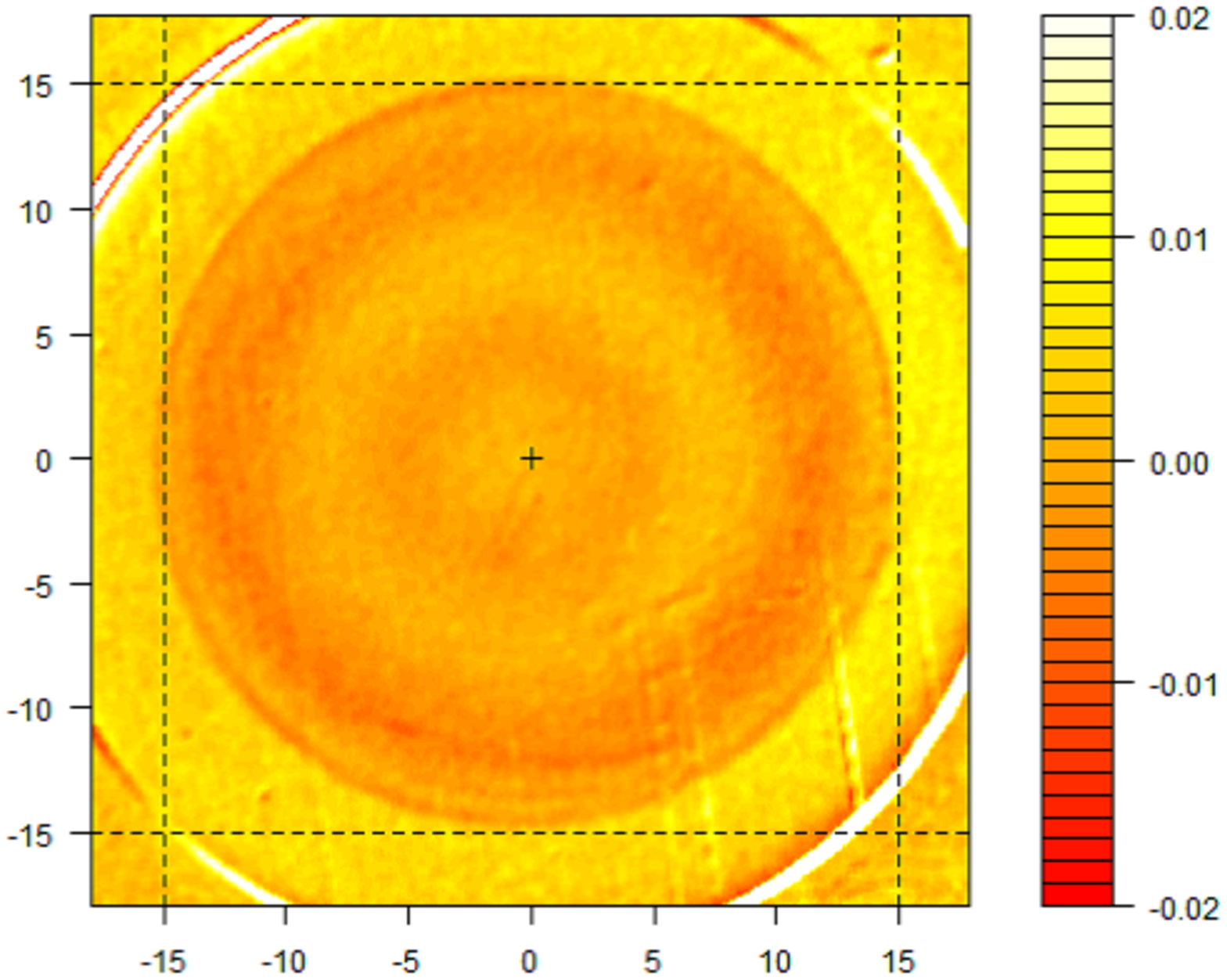}\\
	Fig.2. Contour plort
\end{center}
\end{figure}

\begin{figure}[h]
\begin{center}
	\includegraphics[width=10cm]{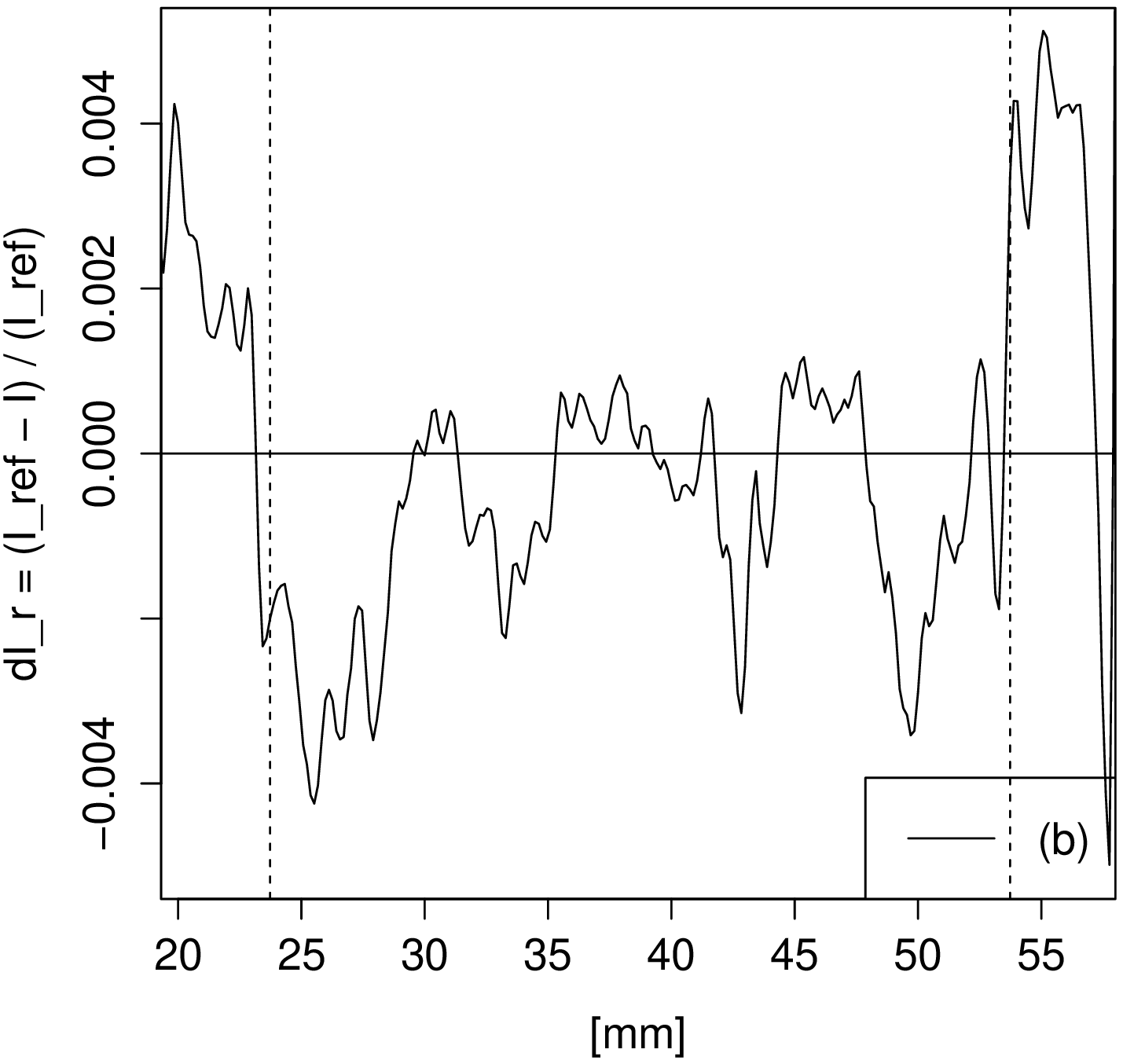}\\
	Fig.3. Cross-section view
\end{center}
\end{figure}
	
	From figure 2, concentric pattern, which is an expected form of pressure distribution caused by cylindrically symmetric sound source, is clearly seen above horn surface. Also the decrease of phosphorescence intensity due to acoustic pressure is evident in figure 3.


	We have developed a new technique for direct measurement of the near-field acoustic pressure distribution based on backside phosphorence measurement of PSP and successfully confirmed its utility experimentally.


\end{document}